\documentclass{aa}             
\usepackage{amssymb,graphicx}  

\begin{document}

\title{Black hole X-ray transients: \\
mass accumulation in the disk - constraints for the viscosity}
\author{E. Meyer-Hofmeister \inst{1} and F. Meyer 
\inst{1}
}
\offprints{Emmi Meyer-Hofmeister\\ e-mail: emm@mpa-garching.mpg.de}
\institute{Max-Planck-Institut f\"ur Astrophysik, Karl
Schwarzschildstr.~1, D-85740 Garching, Germany
} 

\date{Received:s / Accepted:}

\abstract{
The outburst cycles of black hole X-ray transients are now generally
understood as caused by a thermal instability in the accretion disk, the same 
mechanism as in dwarf novae outbursts. During quiescence the accretion occurs
via a cool disk in the outer region but changes to a coronal flow/ADAF
in the inner region. The transition to the coronal flow is caused
by evaporation of matter from the cool disk. This process is an
important feature for the disk evolution.
We point out that if the disk is depleted during the outburst, e.g.
by irradiation, its evolution during quiescence is
independent of the detailed outburst luminosity decline. The mass
accumulation during quiescence has to meet several constraints as the
accretion rate, the recurrence time and the total outburst energy.
We present a critical discussion of different ways to model X-ray
transient outburst cycles and compare with the
requirements from observations. For the case of only little mass left
over after the outburst the observations indicate
a frictional parameter in the cool disk of order $\alpha_{\rm {cold}}$
=0.05, similar to that in dwarf nova disks during quiescence, with no
need to resort to much lower $\alpha$ values of order 0.005.
}
\titlerunning {Disks of black hole X-ray binaries}
\maketitle
%

\section{Introduction}
Soft X-ray transients (SXT), also known as X-ray novae, contain a black
hole or neutron star primary and a low-mass companion star. We
consider here the black hole sources. These systems are
characterized by the high luminosity X-ray outburst after a
long-lasting quiescence and are usually detected in X-rays. An
optical counterpart is known for a number of systems. During
quiescence the optical light is dominated by the secondary star
allowing detailed photometry and spectroscopy to determine the system
parameters of the binary.  A0620-00 was the first low-mass X-ray
binary proven in this manner to contain a black hole (McClintock 
\& Remillard 1986). For this well observed system at the distance of
only about 1kpc two outbursts in optical light were observed,
originally interpreted as classical nova outbursts, 1917 (Nova Mon
1917) and 1975. For several
systems only one outburst is known, their recurrence time is longer
than 30, possibly 50 years. See Tanaka \& Shibazaki (1996) and
Tanaka (2000) for a comprehensive description of the observations of X-ray
transients. Chen et al. (1997)  collected and discussed the available data for
properties of X-ray and optical light curves. Charles (1998) also reviewed
the X-ray, optical, radio and IR observations of black hole X-ray
binaries.

SXTs and dwarf novae are similar in the respect that mass is
transferred from a Roche lobe-filling
low mass secondary star to a compact primary, a neutron star or black
hole in SXTs, a white dwarf in the dwarf nova case. Accretion occurs via a
disk. The evolution of the accretion disk and the triggering of an
outburst therefore were modeled in the same way as for the dwarf nova
outbursts. In the framework of the ``disk instability model'' the
outburst cycles are understood as a thermal relaxation oscillation
between a hot ionized and a cool unionized disk (Meyer\&
Meyer-Hofmeister 1981). Whether the disk is in the cool or the hot
state at certain radii depends on the depends the mass flow rate in
the disk and the amount of mass accumulated, and this on the
viscosity. The size of the disk is important, irradiation can be an
essential fact. The first investigations for SXTs were carried out
by Huang \& Wheeler (1989) and Mineshige \& Wheeler (1989). Later
detailed investigations (Cannizzo et al. 1995, Cannizzo 1998, review
1999, 2000) focussed on modeling the exponential decline of the
outburst lightcurve.

For the evolution of the disk during quiescence one feature is
important in SXTs: it is necessary to have a truncated inner disk.
Otherwise the disk cannot be cool throughout. Even with a very low mass
flow rate the disk would reach high temperatures in the inner region.
This hot region
and the further outward cool region cannot exist together in a
quasi-stationary state. Instead cooling and heating fronts would move
inward and outward in the disk and change the disk back and forth between
the hot and the cool state (Meyer 1981). This truncation of the 
geometrically thin cool disk  naturally results from evaporation of 
mass in the inner disk region into a hot corona
(Meyer \& Meyer-Hofmeister 1994). For the region further in the concept of an
advection-dominated accretion flow (ADAF) was developed as a mode of
accretion which allows to explain the low luminosity of X-ray
transients in quiescence as well as in AGN. The review paper by
Narayan et al. (1998) gives a detailed description of the physics of
ADAFs  as well as the investigations of this mode of accretion by
several authors. The application to galactic black hole X-ray binaries
and galactic nuclei was very successful in modeling the observed spectra.
Best studied systems are A0620-00, V404 Cyg (Narayan et al. 1996,
1997), Nova Muscae (Esin et al. 1997) and GRO J1655-40 (Hameury et al.
1997).

The ADAF in the inner disk region together with the geometrically thin
disk further out provides a consistent picture for the accretion in
SXTs. From observations for the quiescent state 
properties of the accretion disk can be derived. 
The ``standard disk instability model'' for dwarf nova outbursts
allows to understand the triggering of the outbursts
in SXTs after the long recurrence time. This has been proven in the
application to A0620-00 (Meyer-Hofmeister \& Meyer 1999, hereafter
MHM99), using the viscosity values typically found for dwarf novae. There
it was assumed that the disk is practically empty after
the outburst due to the strong irradiation of the disk which prevents the
disk from changing back to the cool state and causes the long lasting
outburst decline (King \& Ritter 1998).
If this is the case the accumulation of mass in the disk is independent
of the details of the outburst. In view of the difficulties which
arise in modeling the observed outburst decline (see e.g. Cannizzo 2000)
it is an advantage to separate the two phases of disk evolution.

Since
the result of King \& Ritter (1998) has important consequences for our
treatment we shortly repeat here their argumentation. They discuss the
optical/X-ray flux ratios already found in earlier investigations and
point out that the optical emission during the outburst of X-ray transients is
far too high to result from intrinsic dissipation only. They argue
that irradiation from the central source is important during the
outburst. The irradiation then prevents the disk from returning to the cool
state until central accretion is greatly reduced. This is consistent
with the long-lasting outburst decline. This is a convincing picture
and it seems difficult to explain the outburst duration without such a
long-lasting accretion and the consequential essential reduction of
disk mass. The analysis of Shabaz et al. (1998) gives further
confirmation for this picture (see Sect. 3.3).  

Menou et al. (2000) model the full outburst cycle
of A620-00 in a way that about 90\%
of the mass in the disk remain there during the outburst. Evaporation
was taken into account in a rough approximation (details will be
discussed later). They conclude that the viscosity in quiescence has
to be very low ($\alpha_{\rm{cold}}$=0.005) to reproduce the long
recurrence times. In their investigation the effect of disk irradiation
during the outburst is not included, also the possibility to model A0620-00
with standard viscosity values as in MHM99 was not noted apparently.  

The aim of the present investigation is a critical comparison of the
predictions of various theoretical models, including earlier results from
Cannizzo (1998), with the available data for black hole transients.

In Sect. 2 we show examples of computed disk evolution for A0620-00.
Evaporation in the inner region reduces the accumulation of mass 
and for low mass transfer rates the instability can only marginally be
triggered. The evaporation of mass therefore
is the important feature which governs the disk evolution during
the long-lasting quiescence and needs to be taken into account for the
evolutionary computations in a way, so that the physics is
included. In Sect. 3 we discuss how far the models agree with the requirements
from observation. In Sect. 4 we focus on the total amount of mass
accumulated when the outburst occurs and compare with collected data
for several black hole SXTs. In Sect. 5 we discuss consequences for
the viscosity in the case of long recurrence time. Additional
computations shed light on the situation in systems with short
recurrence time where an essential part of the mass in the disk might
be left over after the outburst. 

\section{Disk evolution during quiescence - models for A0620-00}
\subsection{Computational method}

Diffusion governs the evolution of the mass distribution in the disk.
Given a black hole mass and an orbital period the mass transfer rate
determines the disk evolution (given an initial mass distribution).
For systems with long recurrence time we assume that the disk
becomes depleted during the long-lasting outburst, caused by
the irradiation which prevents the disk to go back to the cool state
(King \& Ritter 1998). This has an essential consequence for the
conclusions.

Our computer code includes the variation of the outer and inner disk edge
during the evolution. The disk size changes as determined by
redistribution of matter and angular momentum. The size is limited by
the 3:1 resonance radius (Whitehurst 1988, Lubow 1991), which lies inside
the tidal truncation radius for systems with a small mass ratio as the
SXTs. For the initial size of the disk we assume  90\% of the 
resonance radius. The maximal disk size is related to the orbital period and
the secondary star mass.
We take the secondary star mass according to the mass of a Roche-lobe filling
main-sequence star for the given period. The location of the inner disk
edge also varies. It is determined by the evaporation of matter into
the corona. The
total evaporation efficiency computed in a one-zone model is applied
to this interval. For mass flow rates lower than about 0.01 $\dot
M_{\rm{Edd}}$ ($\dot M_{\rm {Edd}}= 40{\pi}GM/{\kappa}c$,
$\kappa$ electron scattering opacity) and a coronal $\alpha$-parameter
of 0.3 the evaporation rate (in g/s) can be
approximated by the following formula
\begin{eqnarray}
\dot{M}_{\rm{evap}}(M,r)& =  & 10^{14.9}
\cdot(M/M_\odot)^{2.3}\cdot(r/10^{9.5})
^{-1.2} 
\end{eqnarray}
with $M$ the mass of the compact star 
(see Liu et al. 1995, Meyer et al. 2000).
Coronal friction like that of a standard disk leads to redistribution of
angular momentum and also to a partial outward flow
that is taken into account (Liu et
al. 1997, application to evaporation around a white dwarf). A detailed
description of the computational method is given in MHM99. 
The black hole mass an important parameter for the disk
evolution, unfortunately can be deduced from the observations only
within some range of uncertainty. We choose the rate of mass
transferred from the secondary star to meet the requirements from
observations. 
** The initial surface density was assumed to be very low, of order of 
the critical surface density ${\Sigma}_{\rm A}$ (viscosity parameter
0.3) or even lower. Without irradiation the
surface density remaining after an outburst usually is between the two
critical surface densities ${\Sigma}_{\rm A}$ and ${\Sigma}_{\rm
B}$. The specific choice does not matter for the
outcome of the simulation, due to the accumulation of much more mass
during the long quiescence until the outburst is triggered.*

\subsection{Requirements from observations}

Given the orbital period (known from observations) and a black hole
mass there are only two free parameters left, the rate of mass transfer from
the companion star and the viscosity parameter in the cool disk 
$\alpha_{\rm{cold}}$. But several requirements have to be fulfilled
connected with quiescence and outburst (this is the
same situation for all X-ray transients). The relevant data with which
the results of modeling should agree are:
\\ (1) The accretion rate in quiescence derived from ADAF model
based spectral fits (Narayan et al. 1998);
\\ (2) the radius of the transition from the outer geometrically
thin disk to the inner ADAF region derived from the Kepler
velocity which corresponds to the maximum
velocity width of the accretion disk H$\alpha$ emission line
(e.g. see Marsh et al.(1994) and Orosz et al. (1994) for A0620-00);
\\ (3) the amount of matter accreted during the outburst
estimated from the outburst light curve (for an assumed distance);
\\ (4) The outburst recurrence time. 

The mass transfer rate affects the outcome for all these requirements. In the
accretion disk the transition to an coronal flow occurs where the
evaporation rate exceeds  the mass flow rate in the disk. The computed
evolution then self-consistently determines the location at each time.
The transition radius becomes smaller with proceeding mass accumulation
and hence increasing mass flow rate in the inner disk
(compare Fig. 3, MHM99). The viscosity value affects the
amount of mass in the disk, higher for lower viscosity.

\subsection{Matter accumulation}

For different models, all designed to describe the X-ray nova
A0620-00, we show in Fig. 1 how the mass in the disk
accumulates with time. The amount is highest in the model of
Cannizzo (1998, Fig. 9, model with long recurrence time).
The viscosity is chosen as proportional to $(h/r)^{1.5}$,
$h$ pressure scale height. Less than 10\% of
the mass in the disk is accreted during the outburst. This is the
scheme used successfully in modeling of dwarf nova outbursts, but probably
not adequate if irradiation affects the mass flow during outburst.
In recent computations of only the outburst (Cannizzo 2000)
irradiation is included and much more mass accretes. In the modeling
of Menou et al. (2000) $\alpha_{\rm{cold}}$=0.005
is taken, irradiation is not included and most of the mass
remains in the disk during the
outburst. (In their investigation the disk evolution was computed with
and without a truncated disk. But for the non-truncated disk and their
further parameters chosen for modeling A0620-00 the recurrence time
is only 5 years, see Fig. 1. We compare with their result for a
truncated disk, Fig. 7). The dashed line gives the total amount of
mass transferred (in Cannizzo's model the total amount of mass
transferred is practically the same as the amount accumulated).

The two lower curves in Fig. 1 (solid lines) show the mass accumulation in our
models for 4 and 6 $M_\odot$ black holes (viscosity in the cool disk
$\alpha_{cold}$=0.05). The mass transfer rates were
chosen to meet the constraints from observations (compare Table 1 in
MHM 1999), $1.9\cdot 10^{-10}M_\odot/yr$ and $3.4\cdot 10^{-10}M_\odot/yr$,
respectively. The
higher amount for the disk around the 6 $M_\odot$ black hole results
from the then higher evaporation rate (see Eq. 1).
This has the consequence that the
critical surface density can only be reached at larger distance $r$
and more mass needs to be stored in
the disk before an outburst occurs (see Fig. 5, MHM99). The dashed line
indicates the total amount of
matter transferred for 4 $M_\odot$. During the late quiescence an
increasing part of the matter transferred is evaporated and accreted, less and
less mass is still accumulating until the critical surface density for
triggering the outburst is reached.

For comparison we show the disk
evolution for a mass transfer rate only 5\% lower (dot-dashed
line  - note that this is not a fluctuation but a difference of the
average over several decades): no outburst occurs,
the disk becomes stationary. Such a system
resembles a case of faint nontransient X-ray binaries.
As discussed in earlier work (Meyer-Hofmeister \& Meyer 2000) the
consideration of such a stationary disk raises
the question how many black hole binaries exist in permanent quiescent
state (see also Menou et al. 1999). Certainly there is a range of
transfer rates around the marginal rate where fluctuations could bring
the system to the other class. Thus the division into the two
classes, stable/unstable, is not sharp. But we do not expect this to
be a model for the SXT outbursts as suggested with the ``globally
stable disks'' where an essential part of all mass accumulated is 
transferred in a rare mass transfer fluctuation, discussed below.

\begin{figure}[ht]
\includegraphics[width=8.8cm]{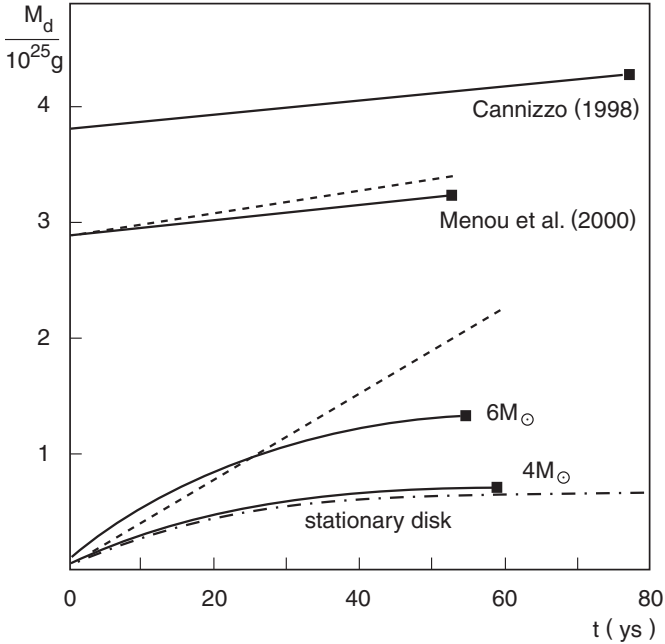}
\caption{Accumulation of mass in the disk during quiescence for
different models, solid lines: Cannizzo (1998), Menou et al. (2000),
Meyer-Hofmeister \& Meyer (1999), 4 and 6$M_\odot$, squares mark the
onset of the outburst; dashed lines: total amount of mass transferred
from the companion star; dashed-dotted line: quasi-stationary disk, see text.}
\end{figure}

Menou et al. (2000) also considered a model of a ``globally stable 
truncated disk'',
that means a disk where the surface density remains below the critical value
for triggering an outburst. For this the mass transfer rate 
$10^{15}$ g/s was taken, one third of the rate used for their ``truncated disk
instability model'' where the outburst is triggered when the critical
surface density is reached. In the globally stable disk model the outburst
is thought to occur due to an increase of the rate of mass transfer
from the secondary star. This means that an essential part of the
mass needed for the onset of the outburst is brought over
during the assumed phase of higher mass transfer.
The authors mention that in such a case the finally accumulated mass
in the disk and the resulting outburst is similar to that in the
truncated disk instability model with a continuous accumulation of
mass (compared before). 

\subsection {**Mass transfer variations*}

Since an increase of the mass transfer rate is sometimes appealed to in
connection with triggering an accretion disk instability we want to
discuss here the timescales involved.

We note that what the recurrence time measures is the average of the
mass transfer rate over the long time of quiescence, e.g. 50
years. This is a long time scale for atmospheres of late
type companion stars. Thermal
timescales of the outer layer relevant for mass transfer and other
oscillation and relaxation times are much shorter. Variations of the
mass transfer rate that might occur on such short timescales are thus averaged
out and do not affect the cycle length. Long-term fluctuations of the
transfer rate, on the other hand, are not achieved easily. It is a
misconception that a fluctuation or change of the
braking rate (e.g. by magnetized wind) results immediately in corresponding
changes of the mass transfer rate as the secondary star
otherwise will over- or under fill its Roche lobe. In order to change
the mass transfer rate by a significant factor one must change the distance
between stellar surface and inner Lagrangian point by a value of the order
of an atmospheric scale height. This is typically of order $10^{-4}$ of the
stellar radius. With orbital evolution timescales of $10^{9}$ years this
gives $10^5$ years before any change of braking leads to the response of
the mass transfer rate. Braking rate fluctuations on shorter timescales are 
smoothed out. For a discussion of this see Ritter (1988).

Observational evidence for a very stable transfer rate is present in the
case of WZ Sge, a dwarf nova system with very long recurrence time,
transferring mass via Roche lobe overflow. That system shows a
repetitive cycle that is observed twice with nearly exactly the same
cycle length of 32.6 and 32.4 years. *

How well do the various models for A0620-00 now meet the requirements
from observations?

\section {Comparison of models for A0620-00 with the observation}
\subsection{Observational requirement (1), evaporation rate - mass
accretion rate}

The accretion rate found from the ADAF model ($M$=6.1$M_\odot$)
fit of the spectrum of A0620-00   
is 3-6$\cdot 10^{15}$g/s (Narayan et al. 1997, 1998). In Table 1 we summarize
the results of the models with respect to the evaporation rate.
To compare with the observations we consider
the theoretical values for 20 years after the outburst that occurred
1975, reduced by the wind loss ($\approx$ 20\%, Meyer et al. 2000).
The theoretical evaporation
rate should roughly agree with the accretion rate derived
from observations.
In Cannizzo's model the assumed evaporation gives values as
low as 2$\cdot 10^{14}$g/s. In the model of Menou et al. (2000) an
assumed evaporation rate
is taken, designed to be significantly smaller than the mass
transfer rate and a parameter value $\varepsilon$ is chosen to ensure
that the maximally truncated disk is still unstable. The value of the
accretion rate after 20 years is 9\,$10^{14}$ g/s, less than one
third of the value derived from the observations.
In our models the evaporation rate (Eq.1) is determined
according to the evaporation theory and is not free. Only the 
parameter for the viscosity in the hot corona enters, chosen as 0.3
in all our computations (comparable to the parameter in the ADAF
modeling and to results of numerical simulations of MHD turbulence
(Hawley et al. 1995).

\begin{table}
\setlength{\textwidth} {8.8cm}
\caption{Mass accretion rates for A0620-00}
\begin{tabular}{lrlr}
\hline
& & & \\
 & BH mass & $\alpha_{\rm cold}$ &${\dot M}_{\rm acc}$ \\
&  $(M_{\odot})$ & &(g/s)\\
\hline
& & & \\
\it{observations}: & & & \\
\rm{spectral fit by Narayan} & & & \\
et al. (1997, 1998)& 6.1 & & 3-6\,$10^{15}$ \\
(data McClintock & & \\
 et al. 1995) & & & \\
\hline
& & & \\
\it{models}: & & & \\
\rm{Cannizzo (1998)} & 10. & $\sim (h/r)^{1.5}$ &2.\,$10^{14}$ \\
Menou et al. (2000) & 6. & 0.005 & 9.\,$10^{14}$ \\
MHM99 & 4. & 0.05 & 3.7\,$10^{15}$\\
MHM99 & 6. & 0.05 & 5.5\,$10^{15}$\\
& & & \\
\hline
\it{stationary disk} & & & \\
\rm{(MHM99)} & 4. & 0.05 & 2.8\,$10^{15}$\\
\hline
\end{tabular}
\vspace {1cm}
\\
Note: Model mass accretion rates taken 20 years into quiescence,
which corresponds after the outburst in 1975 to about the year 1995;
for the models MHM99 the rates of accretion onto the black hole
(evaporation rate minus 20\% wind loss) are given.  
\end{table}

\subsection{Observational requirement (2), transition radius}

For Cannizzo's model no results are given for the inner truncation
of the disk. In the model of Menou et al. the transition radius
follows from their choice of an evaporation formula
and is about $10^{10}$cm 20 years after outburst.
In our treatment of evaporation the transition radius depends on 
the mass flow rate in the disk, the location therefore changes during
the evolution. For 20 years after an outburst the radii found from
our computations for 4 and 6
$M_\odot$ are close to $10^{10}$ cm, in good agreement with
observations. The inner disk edge derived from the maximum
velocity of the H$\alpha$ emission line, $v_{\rm {in}}=2100$km 
(Orosz et al. 1994), is located at 1.3 or 1.1$\cdot 10^{10}$cm
for a 4.4 or 6.1 $M_\odot$ black hole (Narayan et al. 1996, 1997).

\subsection{Observational requirement (3), total amount of mass in the
disk}

Fig. 2 shows the amount of mass $M_{\rm d}$ (total amount of mass in
the cool disk) from model computations for
A0620-00, together with data from observations.  The computations
for 4$M_\odot$ agree better with the total amount of mass derived from
observations.
In earlier investigations (Narayan et al. 1996) a value of 4.4$M_\odot$,
corresponding to an inclination of $70^{\circ}$, was used
to model A0620-00, later Barret et al. (1996) suggested an
inclination of $55^{\circ}$ which leads to a black hole mass of 6.1 $M_\odot$.
The evaporation efficiency still depends on the exact value of the
viscosity used for the hot coronal gas. The result might change
somewhat if another value would be chosen, further investigations are
planned. The results of
Cannizzo (1998) and Menou et al. (2000)
are based on model computations where during the outburst most of the
mass remains in the disk. In Fig. 2 we therefore also give the amount
$\Delta M_{\rm d}$ accreted during the outburst. If irradiation during
the outburst were not important the value $\Delta M_{\rm d}$ of
Menou et al. (2000) would be in agreement with
the estimates from observations. The key assumption of an empty disk
after the outburst, used in our models, leads to a much lower amount of mass.
(In the recent investigation of Cannizzo (2000) irradiation during the
outburst is included and more mass than in the foregoing model is
accreted; but the disk evolution in quiescence is not computed).
In the next section we compare more general results for disks of
different size with observations. 

\subsection{Observational requirement (4), recurrence time}

The models of Cannizzo (1998) and Menou et al. (2000) were developed
to show that long recurrence times can be achieved. Slight changes in the
parameters choosen might reproduce the exact recurrence time, certainly
involving also changes in the mass accumulation. Our
computations yield the observed recurrence time together with all the
other requirements from observations.

\section{Mass accumulation in disks around black holes of different mass}
\subsection{SXTs and WZ Sge stars}

A similarity of SXTs and WZ Sge stars, a subgroup of dwarf novae,
was often pointed out (see e.g. Kuulkers 1999).
This similarity concerns the long recurrence times and
the long duration of the outbursts, the best example being WZ Sge
(Patterson et al. 1981). For a review on the modeling of WZ Sge stars
see Osaki (1995). The total amount of mass 
accreted on to the white dwarf during an outburst (1-2 $\cdot$
$10^{24}$g, Smak 1993) is actually of the same order of magnitude
as the amount of mass involved in SXT outbursts (Chen et al. 1997).

However the  different disk size in these two classes of systems 
has interesting consequences for the
viscosity (in Sect. 5.3 we discuss its possible origin).
The size of the disk in WZ Sge (orbital period 81.3
minutes) is only 1.6$\cdot 10^{10}$cm, limited by the 3:1 resonance
radius. In A0620-00 the disk is roughly ten times larger.
To store the same amount of mass as accumulated in SXT disks in the much
smaller disk of WZ Sge requires a lower value of the
viscosity parameter $\alpha_{cold}$ in WZ Sge systems (0.001
Meyer-Hofmeister et al. 1998). In the larger SXT disks, on the other
hand, the similar amount of accumulated mass points to 
a standard value of a few times 0.01.

\begin{table*}

\pagestyle{empty}
\setlength{\topmargin}{-2.5cm}
\setlength{\textwidth} {17cm}
\setlength{\oddsidemargin}{-2.0cm}
\setlength{\footskip}{-1.0cm}

\caption{X-ray transients established as black--hole sources}

\begin{tabular}{llllllllll}
\hline
\hline
& & & & & & & & &\\
Source & &BH mass & companion & orbital & outburst & rec. time
&distance & $\Delta M_{\rm d}$ & $M_{\rm h}$
\\
name & & & star & period & year & $\Delta t$  &
 &  & \\
& & ($M_\odot$)& & (h)&  & (ys)& (kpc)& $(10^{24}$g) &$(10^{24}$g)\\
\hline
\\
J0422+32 & XNova Per & 3.57$\pm$ 0.3 & M 2 V & 5.1 & '92 & $>$30&2.2&0.8&0.8 \\
A0620-00 &XNova Mon &4.9-10. &K 4 V &7.8 &'17,'75 & 58 & 0.87&3.3&2.5\\
GS2000-25& XNova Vul& 8.5$\pm$ 1.5&K 2 - K7 V &8.3 &'88 &$>$30 &2& 
11.7&2.7 $^*$)\\
GS1124-68&XNova Mus &5.-7.5 &K 2 V &10.4 &'91 &$>$30&4&10.&8.8  \\
H1705-25&XNova Oph7 &4.9$\pm$ 1.3 &K3 V&12.5 &'77  &$>$30 &4.3&3.3&- \\
4U1543-47&XN '71,'83,'92 &2.7-7.5 &A 2 V & 27.0&'71,'83,'92  &$\approx 10$
 &4&8.1&18.3$^*$) \\
J1655-40&XNova Sco &7.02$\pm $0.22 &F 3-6 IV &62.7 &'94 &$>$30
& 3.2&0.5& 4.2\\
GS2023-338&XNova Cyg &12.3$\pm$ 0.3 &K 0 IV &155.3 &'38,'56,'79,'89&10-20 
&3.5&58.9&-\\
& & & & & & & & &\\
\hline
\end{tabular}
\vspace {1cm}
\\
Note: 
Data for black hole mass, spectral type of companion star, orbital period, 
outburst year, distance and  mass accreted during outburst $\Delta
M_{\rm d}$ from Chen et al. (1997, see references therein); given in
addition is $M_{\rm h}$, the  amount of mass in the disk heated
at the start of the outburst $M_{\rm h}$, estimated by Shabaz
et al. (1998) (see text).

Data from other sources:  distance for
GS 1124-68 as used by Shabaz et al. (1998), $\Delta M_{\rm d}$ according
to this value; companion
star spectral type for 4U 1543-47 from Tanaka (2000). $^*$)
the difference between $\Delta M_{\rm d}$ and $M_{\rm h}$ arises from
a discrepancy in the value $\Delta M_{\rm d}$ for GS 2000+25,
and from the larger distance, 8 kpc, for 4U1543-47
(Shabaz et al. 1998) respectively.
\\
\end{table*}  

\subsection{Data from observations of SXTs}

In Table 2 we give data for systems established as black hole
transient sources. Knowledge of the distance of the sources is
important for the estimate of the luminosity and therefore also the
outburst energy and amount of
accreted mass. Unfortunately for some systems, e.g. GS 1124-68,
the values are very uncertain. For J0422+32 recently the distance 
1.39 kpc was derived (Webb et al. 2000). The properties of black hole
transients
compiled in the reviews of Tanaka and Shabazaki (1996), Chen et
al. (1997), Charles (1998) and Tanaka (2000) mainly differ with
respect to the black hole mass estimates.

For our modeling the black hole masses
are important, since the evaporation efficiency depends on it. 
Chen et al. (1997) derived estimates for the amount of mass accreted during the
outburst from the outburst light curves of the systems.
In the analysis of Shabaz et al. (1998) the exponential/linear decays
of the light curves were studied and estimates for the amount of
mass heated during the outburst were derived.

\begin{figure}[ht]
\includegraphics[width=8.8cm]{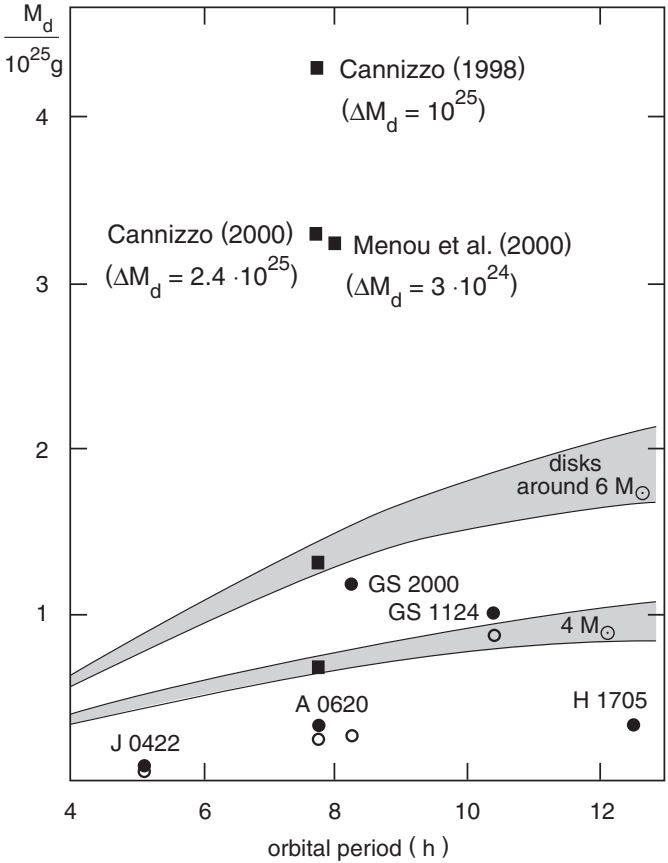}
\caption{Comparison of the total amount of mass accumulated in the
disk for the outburst, in theoretical models and derived from observations..
Models: shaded areas: results for disks of different size,
for an appropriate range of mass transfer rates (discussed in the
text). 
Observations: filled circles: values from observed outburst lightcurves
(from Chen et al. 1997), open circles:
estimates for the amount of heated mass in the disk 
(from Shabaz et al. 1998), compare Table 2. }. 
\end{figure}

\subsection {Comparison with observations}

In Fig. 2 we show more general results for mass accumulation in disks
around 4 and 6$M_\odot$ black holes. The disk size depends on the orbital
period. We computed the disk evolution for 
periods from 4 to 12 hours and different mass transfer rates. The
range of rates was taken as follows: the
minimal rate arises from the fact that for too low rates no outburst
occurs anymore (the higher the rate the more often outbursts are
triggered), we took the maximal rate as the one which initiates
outbursts every ten years. We did not want to consider shorter
recurrence time cycles for our analysis of transients. The
shaded areas in Fig. 2 mark the range of values of total
mass accumulated corresponding to this range of transfer rates. Our two
models for A0620-00 lie in these ranges. 
Later in this paper we compare with data for a number of SXTs
observations. 

In a very detailed investigation of outburst light curves of SXTs Chen
et al.(1997) derived estimates for the amount of mass accreted. 
We show these values in Fig. 2 (for GS 1124-68 we have taken the smaller
distance 4 kpc as used in Shabaz et al. 1998). These values are
uncertain  if the distance is not well known. In their investigation
of the effect of irradiation of the accretion disk during the outburst
Shabaz et al. (1998) give estimates for the mass in the disk heated in
the outburst. **The question is whether this amount of mass includes all
mass present in the disk at the time an outburst is triggered. The authors
found that the radius of the hot disk at the peak of the outburst is
comparable to the circularization radius. For A0620-00 they give the
value $0.6\, 10^{11}$ cm. Therefore the main part is indeed hot.
We give the values of heated mass in Fig. 2. All estimates for
the amount of mass accreted in the outburst or heated are in the range
$10^{24}$ to 1.2$\cdot 10^{25}$g.*

\section{Viscosity in the cold disk}
\subsection{Systems with long recurrence time}

We first discuss the situation in systems with long recurrence time
and long-lasting outbursts. If in these systems the
irradiation causes the long-lasting decay (King \& Ritter 1998)
the disk is depleted. The triggering of the outbursts after decades can well
be modelled with a standard value around 0.05 as usual for dwarf nova
outburst computations (all the observational requirements are fulfilled,
see Sect. 3 and earlier work (MHM99)). Inspite of this result
Menou et al. (2000) claimed that the viscosity would have to be
low ($\approx$ 0.005) to model the SXT outbursts. Disk
structure calculations show that the surface density and therefore the
mass in the disk scales with $\alpha ^{-0.8}$ approximately. For a
viscosity ten times lower the amount of mass accumulated and the 
outburst energy would be roughly ten times higher. In the modeling with
low viscosity the way out always is that only a small part of the total
mass accretes. This is in conflict with the effect of irradiation and
in general disagrees with the above mentioned amounts of mass heated
up in the disk.

\subsection{Systems with short recurrence time}

There are a few systems with  more frequent outbursts. 4U 1630-472 e.g. has
outbursts about every two years. Their duration differs, also the observed
luminosity which reaches values around 4$\cdot 10^{38}$ erg/s. The amount of
mass accumulated in the disk is derived as several
$10^{25}$g, based on the uncertain distance of 10 kpc (values from
Chen et al. 1997). It is unclear whether the disk is empty after the
outburst.

In an earlier investigation such short cycles were already computed 
assuming a depleted disk and  $\alpha _{\rm{cold}}$=0.05 
(Meyer-Hofmeister \& Meyer 2000). Now we
performed test computations to clarify whether, in principle, these
short outburst cycles can also be modeled with a not depleted disk and the
standard viscosity value $\alpha _{\rm{cold}}$=0.05. The black
hole mass and the orbital period of 4U 1630-472 
are not known. We performed test calculations for 6$M_\odot$ and 7.75
hours (period as A0620-00). For the initial mass distribution we assume a
surface density distribution as adequate from dwarf nova outburst
modeling experience (log$\Sigma_{\rm{initial}}(r)$=
0.5(log$\Sigma_{\rm A}(r)$+log$\Sigma_{\rm B}(r)$), $\Sigma_{\rm A}$
and $\Sigma_{\rm B}$ critical surface density values), viscosity in
the hot state 0.3).
The mass transfer rate 2\,$10^{-9}M_\odot$/y is high enough to produce
the outburst after  two years. The accumulated amount of mass in the disk
is 1.1$\cdot 10^{25}$g, in rough agreement with the observations. We conclude
that values of order 0.05 are also suitable to describe the short outbursts in
cases where the disks are not depleted.

\subsection{The general picture of the origin of the viscosity in
cold SXT disks during quiescence}

The values, $\alpha_{\rm{cold}}\approx$0.05, in quiescent X-ray transients
appear to support a general picture of magnetic
friction in disks in SXTs and dwarf novae. In outburst the magnetic
field caused viscosity (Balbus \& Hawley 1991, 1992) is high sustained by
dynamo action in the accretion disk (for recent numerical simulations
see Hawley et al. 1995, Brandenburg et al. 1995, Armitage 1998, Hawley
2000).
As the disk returns to the quiescent state and becomes cool the
electrical conductivity drops and the dynamo can no longer operate 
(Gammie \& Menou 1998), the high outburst viscosity vanishes. However a
magnetosphere of the secondary star in which the accretion disk is
embedded can still provide magnetic flux through the disk on which the
Balbus-Hawley mechanism can work and supply a smaller magnetic friction in
quiescence than that in outburst. WZ Sagittae systems have
probably lost this source of magnetic flux since their secondaries have
become cool degenerate brown dwarfs which
can not sustain dynamo action or keep fossil fields in their interior.
This could explain why their remaining viscosity is so much smaller
than that of all other systems in quiescence though they have similar
viscosity in outburst (Meyer \& Meyer-Hofmeister 1999). Then the
derived $\alpha$-values for SXTs in
quiescence would indicate that their secondaries have magnetospheres.
In this respect we also note that these stars appear to be
evolved (King et al. 1996) and tend to have outer convective zones, an
ingredient for solar type dynamos.

\section{Conclusions}

King \& Ritter (1998) argued that strong irradiation during the
long-lasting outbursts of SXTs keeps the disk hot, and almost all mass
in the disk
flows towards the black hole. We make the key assumption that in black
hole transients with long recurrence time the disk is significantly depleted
at the end of the outburst. The estimates
for the amount of mass in the disk heated during the outburst
(Shabaz et al. 1998) confirm this picture. When one follows the disk
evolution and mass
accumulation one finds that the long recurrence times can well be modelled
with a  ``standard'' viscosity value $\alpha_{\rm{cold}}$=0.05.
Evaporation of the inner disk region is an important feature in such
disk evolution. All constraints from observation are fulfilled.
For a full outburst cycle, where the  disk evolution during quiescence
computed here would be completed with the computation of the run of
the outburst follows: the amount of matter accumulated agrees
with the estimates from the outburst lightcurve. Only the features of
the detailed lightcurve are not determined, those depending on
parameters as the viscosity in the hot state and irradiation.
We point out that the evaporation efficiency is taken from a physical
model (Meyer \& Meyer-Hofmeister 1994) and no ad hoc parameter fitting
was used. The computations of Menou et
al. (2000) are based on a simulation of evaporation with free parameters
which ensure that the evaporation rate is significantly below the mass
transfer rate and, more severe, the maximally truncated disk is still 
unstable. Such a strategy does not allow a discrimination between
stable and unstable models.

The models of
Cannizzo (1998, $\alpha$ proportional to
$(h/r)^{1.5}$, $h$ pressure scaleheight) and Menou et al. (2000,
$\alpha_{\rm{cold}}$=0.005), suggest that only a small part of the
total mass in the disk is accreted during an outburst. 
But, even if the disk in A0620-00 would not be depleted the models
only agree marginally with the other requirements
from observations.

We further find that SXTs with a recurrence time as short as two years
can also be modeled with $\alpha_{\rm{cold}}$=0.05, assuming a
depleted or a not depleted disk.

We thus conclude that the accretion disk in SXTs in quiescence display
rather standard cold disk friction common to SXTs and dwarf novae and
do not need nor indicate very low $\alpha$-values of order 0.005.

\end{document}